\documentclass[12pt]{article}

 \newcommand{\bea}{\begin{eqnarray}}
\newcommand{\eea}{\end{eqnarray}}

 \def\m{{\mu}}
 
 \def\n{{\nu}}
 \def\ep{{\epsilon}}

 \def\frac#1#2{{#1\over #2}}

 \def\s{\sqrt}

\def\be{\begin{equation}}
\def\ee{\end{equation}}
\def\ba{\begin{eqnarray}}
\def\ea{\end{eqnarray}}
 \def\ep{\epsilon}

 \def\de{\partial}

 \def\ti{\tilde}
 \def\ap{\alpha}

 \def\no{\nonumber \\}

 \def\la{\langle}
 \def\lb{\rangle}
 \def\ep{\epsilon}

\newcommand{\ex}[1]{\mathrm{e}^{#1}}

\newcommand{\pa}[1]{\left(#1 \right)}

\newcommand{\BR}[1]{\Biggl[#1 \Biggr]}

\newcommand{\bb}[1]{\mathbb{#1}}

\newcommand{\ca}[1]{\mathcal{#1}}

\newcommand{\abs}[1]{\left|#1\right|}
\newcommand{\ave}[1]{\langle #1\rangle}
\newcommand{\subs}[2]{\renewcommand{\arraystretch}{0.5}\begin{array}{c}\scriptstyle #1 \\ \scriptstyle #2\end{array}\renewcommand{\arraystretch}{1}}

\newcommand{\fr}{\frac}

\def\m{{\mu}}
 
 \def\n{{\nu}}
 \def\ep{{\epsilon}}

 \def\e{{\epsilon}}

\def\ii{{\mathrm{i}}}

\usepackage[hypertex]{hyperref}
\usepackage{url}
\usepackage{color}
\usepackage{graphicx}
\usepackage{dcolumn}
\usepackage{amsmath,amssymb}
\usepackage{bm}

\begin{document}

\begin{titlepage}
\thispagestyle{empty}

\begin{flushright}
YITP-17-126
\\
IPMU17-0163
\\
\end{flushright}

\bigskip

\begin{center}
\noindent{{\large \textbf{Renyi Entropy for Local Quenches in 2D CFTs\\
from Numerical Conformal Blocks}}}\\
\vspace{2cm}
Yuya Kusuki$^{a}$ and Tadashi Takayanagi$^{a,b}$
\vspace{1cm}

{\it
$^{a}$Center for Gravitational Physics, \\
Yukawa Institute for Theoretical Physics (YITP), Kyoto University, \\
Kitashirakawa Oiwakecho, Sakyo-ku, Kyoto 606-8502, Japan\\ \vspace{3mm}
$^{b}$Kavli Institute for the Physics and Mathematics of the Universe,\\
University of Tokyo, Kashiwano-ha, Kashiwa, Chiba 277-8582, Japan\\
}
\vskip 2em
\end{center}

\begin{abstract}
We study the time evolution of Renyi entanglement entropy for locally excited states in
two dimensional large central charge CFTs. It generically shows a logarithmical growth and we
compute the coefficient of $\log t$ term. Our analysis covers the entire parameter regions
with respect to the replica number $n$ and the conformal dimension $h_O$ of the primary operator which creates the excitation. We numerically analyse relevant vacuum
conformal blocks by using Zamolodchikov's recursion relation. We find that the behavior of the conformal blocks
in two dimensional CFTs with a central charge $c$, drastically changes when the dimensions of external primary states reach the value $c/32$. In particular, when $h_O\geq c/32$ and $n\geq 2$, we find a new universal formula $\Delta S^{(n)}_A\simeq \frac{nc}{24(n-1)}\log t$. Our numerical results also confirm existing analytical results using the HHLL approximation.

\end{abstract}

\end{titlepage}


\section{Introduction and Summary}

One useful way to characterize dynamical aspects of quantum field theories (QFTs) is to
study properties of entanglement entropy \cite{BKLS,Sr,HLW,CC,FSL,FSLL,FSLLL}. Especially, the evolutions of entanglement entropy when we excite QFTs provide us with important pieces of information such as whether the quantum field theory is integrable or chaotic.
Indeed, this has been manifest for locally excited states in conformally field theories
(CFTs) as we briefly review below.

 A famous example in this direction is the studies of global quantum quenches, which are translationally invariant excited states created by a sudden change of the Hamiltonian \cite{CCG} (also refer to references in \cite{Qreview}). The time evolution of entanglement entropy under global quenches shows a causal and relativistic propagation of elementary excitations in generic CFTs.
One way to study how the details of propagations of quantum entanglement
are different beyond this universal behavior is to look at local excitations
rather than global excitations. Motivated by this, the main purpose of this paper is to
study the time evolution of entanglement when we locally excite a class of CFTs which are strongly interacting and have many degrees of freedom. Such a class of CFTs have
holographic duals and is called holographic CFTs \cite{He,Ha,HKS}.

 A locally excited state $|\Psi\lb$ (operator local quench) is defined by acting with a local operator $O(x)$ on the CFT vacuum $|0\lb$ in the manner\footnote{We would like to stress that
$\ep$ in (\ref{lopw}) is the UV cut off of the local excitations and should be distinguished  from the UV cut off (i.e. the lattice spacing) of the CFT itself.}
\be
|\Psi\lb={\cal N}e^{-\ep H}O(x)|0\lb, \label{lopw}
\ee
where ${\cal N}$ is the normalization factor.
 The infinitesimally small parameter $\ep>0$ provides a UV regularization as the truly localized operator has infinite energy.
 Consider the time evolution of the entanglement entropy $S_A=-\mbox{Tr}\rho_A\log\rho_A$ and
more generally Renyi entanglement entropy $S^{(n)}_A=\frac{1}{1-n}\mbox{Tr}\log(\rho_A)^n$
 for the time evolved excited state $|\Psi(t)\lb=e^{-iHt}|\Psi\lb$. We choose the subsystem $A$ to be the half-space and $\rho_A$ is the corresponding reduced density matrix. The excitation is originally located in the subsystem $B$ (i.e. complement of $A$), thus it creates additional entanglement between them. The main quantity of interest is the growth of  entanglement entropy  compared to the vacuum:
 \be
 \Delta S^{(n)}_A(t)=S^{(n)}_A(|\Psi(t)\lb)-S^{(n)}_A(|0\lb).\label{difs}
 \ee
Note that the $n=1$ limit coincides with the von-Neumann entropy (or entanglement entropy)
growth $\lim_{n \to 1}\Delta S^{(n)}_A(t)=\Delta S_A(t)$. If we define $l$ to be the distance between $x$ and the boundary point $\de A=\de B$, $\Delta S^{(n)}_A(t)=0$ for $t\leq l$ as follows from the
causality. For $t>l$, $\Delta S^{(n)}_A(t)$ gets non-vanishing in general as the excitations in the region $B$ can reach the region $A$.

 Calculations of $\Delta S^{(n)}_A$ for massless scalar fields have been performed in \cite{Nozaki:2014hna,Nozaki:2014uaa,Nozaki:2015mca,Nozaki:2016mcy} and it was found that the growth $\Delta S^{(n)}_A(t)$ approaches a finite positive constant at late time. This is clearly interpreted as a system of entangled particles propagating at the speed of light \cite{Nozaki:2017hby}. The same behavior has also been found for rational CFTs in two dimensions, which is a typical example of integrable CFTs, \cite{Nozaki:2014uaa,He:2014mwa,Chen:2015usa,Caputa:2015tua,Caputa:2016yzn,Numasawa:2016kmo}.

Furthermore, a recent study of 1+1-dimensional orbifold CFTs found an exotic time evolution $\Delta S^{(n)}_A\propto \log(\log t)$ for irrational (but exactly solvable) CFTs \cite{Caputa:2017tju}. For other field theoretic progress on local quenches refer also to \cite{Shiba:2014uia,Caputa:2014eta,deBoer:2014sna,Guo:2015uwa,Caputa:2015waa,HaSC,Rangamani:2015agy,David:2016pzn,
Sivaramakrishnan:2016qdv,Numasawa:2016emc,He:2017lrg}.

On the other hand, for holographic CFTs, which are strongly interacting CFTs with large central charges \cite{He,Ha,HKS}, the evolution behavior changes drastically. In the calculation
using the holographic formula \cite{RT,HRT}, the local excitation corresponds to a massive particle falling in AdS$_3$, whose mass $m$ is related to the conformal dimension $\Delta_O$ of $O(x)$ in (\ref{lopw}) via the standard relation $\Delta\simeq mR$ ($R$ is the AdS radius). The holographic results for the (von-Neumann) entanglement entropy under time evolution at late time was obtained in \cite{Nozaki:2013wia} and this reads
\be
\Delta S_A\simeq \frac{c_{CFT}}{6}\log\fr{t}{\e}, \label{holee}
\ee
where $c_{CFT}$ is the central charge of the 2D holographic CFT.

This time dependence has been precisely reproduced in \cite{Asplund:2014coa} using a large central charge CFT analysis.
Such a behavior is expected to stem from the chaotic nature of holographic CFTs, where the quasi-particle picture breaks down. Similar calculations in higher dimensional holographic CFTs have recently performed in \cite{Jahn:2017xsg} for a holographic computation in the AdS$_4/$CFT$_3$ setup, where $\log t$ like behavior was observed.

In summary, the functional form of the growth of von-Neumann entanglement entropy $\Delta S_A
=\Delta S^{(1)}_A$  depends on the nature of CFTs roughly in such a way that as interactions in the CFT get stronger, the rate of its growth increases. In \cite{Caputa:2017tju}, it was even conjectured that the logarithmic growth (\ref{holee}) found in the holographic 2d CFTs give an upper bound of the growth rate for any two dimensional CFTs. In this way, we already
had enough knowledge about the von-Neumann entanglement entropy for the locally excited states.

However, to capture more information of quantum entanglement, we need to look at the Renyi entanglement entropies $\Delta S^{(n)}_A$ for various $n$. At present, we have only limited understanding of $\Delta S^{(n)}_A$ for the locally excited states as we will briefly review below. To obtain the full understanding of the time evolutions of the Renyi entanglement entropies is the main purpose of the present paper.
 
For the Renyi entanglement entropy in 2D holographic CFTs, the following behavior was derived in \cite{Caputa:2014vaa} when the (chiral) conformal dimension $h_O$ of local operator is small enough:
 $\Delta_O=2h_O\ll c_{CFT}$:
 \ba
 \Delta S^{(n)}_A\simeq \frac{2nh_O}{n-1}\log\fr{t}{\e}. \label{capee}
 \ea

In this way, interestingly, there are varieties of behaviors of (Renyi) entanglement entropy
for the locally excited states defined by (\ref{lopw}), depending on how much a given CFT is chaotic. Notice that the above operator local quench makes a significant contrast with the original class of local quenches introduced in \cite{CCL} (see also \cite{SDu,Qreview}) defined by joining two semi-infinite CFTs, which always lead to logarithmic
growth of entanglement entropy $\Delta S_A(t)\simeq \frac{c_{CFT}}{3}\log\frac{t}{\ep}$ for any two dimensional CFTs, both integrable and chaotic.

In this paper, we would like to explore more on the analysis of $\Delta S^{(n)}_A$ for 2D
holographic CFTs to obtain complete and systematic understandings without relying on special
approximations. For example, one may notice that it is
not immediately clear why the two results (\ref{holee}) for the von-Neumann entropy and (\ref{capee}) for the Renyi entropy with $h_O\ll c_{CFT}$ are related to each other. Indeed, the latter gets divergent in the von-Neumann limit $n\to 1$. Moreover, there is no known result for
the evolution of Renyi entropy with large $h_O$.

To find a full control of the computability for any values of $n$ and $h_O$ we will
employ the powerful numerical program recently developed in \cite{Kaplan} by Chen, Hussong, Kaplan and Yi, based on Zamolodchikov's recursion relation \cite{Zam, Zamolodchikov:1985ie, Zam2, Chang:2014jta}. For example, this relation was applied to computations of entanglement entropy for multi-intervals in \cite{GR}. This powerful and non-perturbative method allows us to evaluate any conformal blocks with any values of conformal dimensions and central charges $c$. Note that for the replica computation of
the Renyi entanglement entropy $\Delta S^{(n)}_A$ we set $c=n\cdot c_{CFT}$. Therefore in this paper we express the total (replicated) central charge as $c$ when we talk about the conformal blocks.

\begin{figure}[h]
\begin{center}
  \hspace*{-4em}
  \includegraphics[width=10cm]{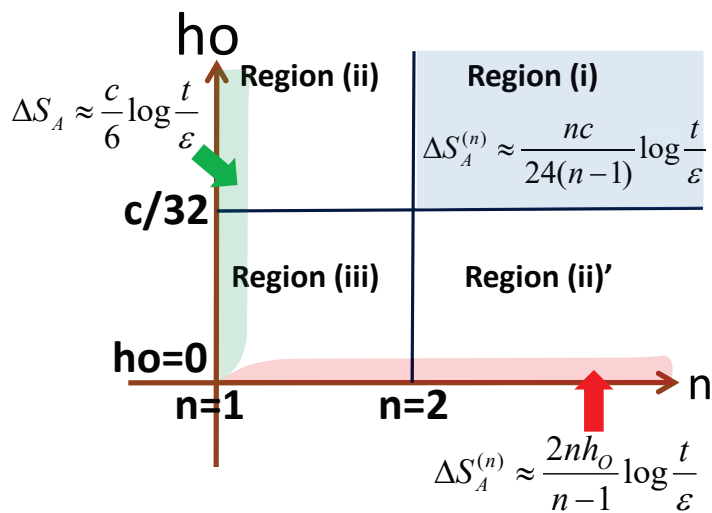}
\end{center}
 \caption{We summarized the behaviors of the logarithmic growth of Renyi entanglement entropy
 $\Delta S^{(n)}_A(t)$. The green and red regions can be well approximated by the HHLL conformal block approximation.}
 \label{fig:phase2}
\end{figure}

As we will explain later in the present paper, our analysis reveals that the behavior of conformal block changes drastically when the conformal
dimensions are large. We find that the behavior of vacuum conformal block
for the 4-pt function of the form $\la O_AO_AO_BO_B\lb$ can be
classified into three regions $(i) h_A,h_B\geq \frac{c}{32}$, $(ii) h_A\leq \frac{c}{32},\ h_B\geq \frac{c}{32}$, and $(iii) h_A,h_B\leq \frac{c}{32}$. The region $(ii)$ is equivalent to $(ii)'\ h_A\geq \frac{c}{32},\ h_B\leq \frac{c}{32}$.
In region $(i)$ the coefficients $c_n$ of the power series of the function $H(q)=1+\sum_{n=1}^\infty c_n q^{2n}$, which is an important part of the contributions to the conformal block, grows polynomially. On the other hand in the region $(ii)$ and $(iii)$, $c_n$ grows exponentially.

In terms of our Renyi entanglement entropy calculations, these regions correspond to
$(i)\ h_O\geq \frac{c_{CFT}}{32},\ n\geq 2$, $(ii)\ h_O\geq\frac{c_{CFT}}{32},\  n\leq 2$,
$(ii)'\  h_O\leq \frac{c_{CFT}}{32},\ n\geq 2$, and $(iii) \ h_O\leq\frac{c_{CFT}}{32},\  n\leq 2$.
In particular, studies of the region $(i)$ lead to the following new universal behavior of Renyi entropy growth
\ba
 \Delta S^{(n)}_A\simeq \frac{nc_{CFT}}{24(n-1)}\log\fr{t}{\e}. \label{kee}
 \ea
On the other hand, in the region
$(ii)$ and $(iii)$, we can apply  the formula (\ref{holee}) when $n\simeq 1$, while in the region $(ii)'$ and $(iii)$ we can apply the formula (\ref{capee}) when $h_O\ll c_{CFT}$. These behaviors are summarized in Fig.\ref{fig:phase2}. It will be an intriguing future problem to reproduce the new behavior (\ref{kee}) from holographic calculations. Also it is an important
to understand better the nature of the non-trivial transition of the conformal blocks at $h_{A,B}=c/32$.

This paper is organized as follows: In section two, we review the computation of Renyi
entanglement entropy in large central charge CFTs. In section three we analyse the vacuum conformal blocks
by using the Zamolodchikov's recursion relation for various parameter regions. In section four, we apply the results of section three to the computations of Renyi entanglement entropy.
In appendix A we summarize our conventions of four point functions and conformal blocks.
In appendix B we briefly review  the Zamolodchikov's recursion relation .

\section{Renyi Entropy and Conformal Blocks}

Here we review the general calculation of Renyi entanglement entropy for excited states in
a 2D CFT in terms of conformal blocks. We express the 2D CFT we consider as $\ca{M}$ and its central charge is written as $c_{CFT}$.

\subsection{Growth of Renyi Entropy for Excited States}

\begin{figure}[h]
 \begin{center}
  \includegraphics[width=6.0cm]{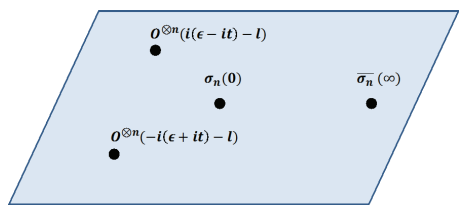}
 \end{center}
 \caption{The positions of operators in the replica computation (\ref{eq:defREE}). }
 \label{fig:pos}
\end{figure}

In the replica computation for Renyi entropy we introduce the replicated CFT with the central charge $nc$. The growth of Renyi entropy (\ref{difs}) for locally excite states (\ref{lopw}) can be expressed by \cite{ABS,Asplund:2014coa}
\begin{equation}
\Delta S^{(n)}_A=\frac{1}{1-n}\log \frac{\ave{O^{\otimes n}O^{\otimes n}\sigma_n \bar{\sigma_n} }}{\ave{O^{\otimes n}O^{\otimes n}}\ave{\sigma_n \bar{\sigma_n}}}.
\label{eq:defREE}
\end{equation}
we take the subsystem $A$ to be a semi-infinite interval and the twist operators are inserted at both end points of $A$. Here $O^{\otimes n}$ is defined on the cyclic orbfold CFT (replicated CFT) $\ca{M}^n/\bb{Z}_n$ (with the central charge $c=n\cdot c_{CFT}$), using the operators in the seed CFT $\ca{M}$ (with the central charge $c_{CFT}$) as
\begin{equation}
O^{\otimes n} = O \otimes O \otimes \cdots \otimes O,
\end{equation}
which is separated by the distance $l$ from the boundary of $A$ as shown in Fig.\ref{fig:pos}.
We take the subsystem $A$ to be the half line $[0,\infty]$. In the complex plane $(z,\bar{z})$, we insert the two twist operators each at $z=\bar{z}=0$ and $z=\bar{z}=\infty$. The replicated operator $O^{\otimes n}$, which produces the state $|\Psi\lb$ of the local excitation, is inserted at $z=i(\ep-it)-l, \bar{z}=-i(\ep-it)-l$ and the other, which creates $\la \Psi|$, is at $z=-i(\ep+it)-l, \bar{z}=i(\ep+it)-l$ as in  Fig.\ref{fig:pos}. The rule of the analytical
continuation from the Euclidean time to Lorentzian time follows from the prescription in 
\cite{CCG}. Now, these insertions of four 
operators lead to the growth of the trace of the reduced density matrix $\mbox{Tr}(\rho_A)^n$ i.e. that of Renyi entropy as in (\ref{eq:defREE}).

We define the chiral conformal dimension of $O(x)$ as $h_O$. Then the dimension of
$O^n$ is written as $nh_O$. The dimension of twist operator $\sigma_n$ is given by the standard formula
\be
h_{\sigma_n}=\frac{c_{CFT}}{24}\pa{n-\fr{1}{n}}=\frac{c}{24}\pa{1-\fr{1}{n^2}}. \label{sgdim}
\ee

By using the cross ratio $z=\frac{z_{12}z_{34}}{z_{13}z_{24}}$, we can rewrite (\ref{eq:defREE}) as
\begin{equation}
\frac{\ave{O^{\otimes n}O^{\otimes n}\sigma_n \bar{\sigma_n} }}{\ave{O^{\otimes n}O^{\otimes n}}\ave{\sigma_n \bar{\sigma_n}}}=\abs{z^{2h_{\sigma_n}}}^2G(z,\bar{z}),
\end{equation}
where $G(z,\bar{z})$ is the four point function (refer to appendix A for more details of
our conventions)
\be
G(z,\bar{z})=\la \sigma_n(0)\bar{\sigma}_n(z)O^{\otimes n}(1)O^{\otimes n}(\infty)\lb,
\ee
and the cross ratio $(z,\bar{z})$ is explicitly expressed as
\begin{equation}
z=\frac{2i \e}{l-t+i \e}, ~~~~~\bar{z}=-\frac{2i\e}{l+t-i\e}.
\end{equation}

\begin{figure}[h]
 \begin{center}
  \includegraphics[width=7cm]{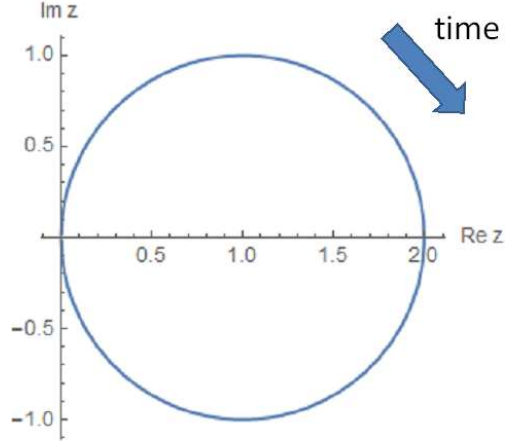}
 \end{center}
 \caption{The time evolution of cross ratio $z$.}
 \label{fig:mon}
\end{figure}

From these expressions, one find that the sign of the imaginary part of the cross ratio $z$ changes at $t=l$.  As a result, the cross ratio $z$ picks up the factor $\ex{-2\pi i}$ at $t=l$ as $1-z \to \ex{-2\pi i}(1-z)$ (see Fig.\ref{fig:mon}). This does not happen for the anti-chiral coordinate $\bar{z}$.
We write the value of $z$ after this monodromy transformation as $z_{mo}$. For example, if $f(z):=\log(1-z)$, then $f(z_{mo})=-2\pi i +f(z)$. In the late time
region $\ep\ll l\ll t$, which we are interested in, we find
 \begin{equation}
z=-\frac{2i \e}{t}\equiv z_{mo}, ~~~~~\bar{z}\simeq -\frac{2i\e}{t}\equiv \bar{z}_{mo}.
\label{monoz}
\end{equation}

Therefore the time evolution of excited Renyi entropy at late time can  be calculated as
\ba
\Delta S_A^{(n)}=\frac{1}{1-n}\log\BR{ |z_{mo}|^{4h_{\sigma_n}} G(z_{mo},\bar{z}_{mo})}.\no
\label{eq:fsa}
\ea

The 4-pt function $G$ can be expressed as a summation over conformal blocks $F^{h_{\sigma_n},h_{O^{\otimes n}}}_{h_p}(z)$. In holographic CFTs we expect that due to
its sparse spectrum, we can approximate $G$ by restricting to the vacuum conformal block
$h_P=0$. Thus we obtain
\ba
\Delta S_A^{(n)}
&=&\frac{1}{1-n}\log\BR{ |z_{mo}|^{4h_{\sigma_n}} \sum_p C_{O^{\otimes n} O^{\otimes n} }^p  C_{\sigma_n \sigma_n }^p |F^{h_{\sigma_n},h_{O^{\otimes n}}}_{h_p}(z_{mo})|^2}\no
&\simeq&\frac{1}{1-n}\log\BR{ |z_{mo}|^{4h_{\sigma_n}}
|F^{h_{\sigma_n},h_{O^{\otimes n}}}_{0}(z_{mo})|^2},
\label{eq:Fform}
\ea
This is the main target which we would like to evaluate in this paper.

\section{Analysis of Conformal Blocks}

In this section, after we review the numerical computation of conformal block
based on Zamolodchikov's recursion relation, we explore properties of conformal block towards
the calculation of Renyi entropy. We are especially interested in large central charge
limit $c\gg 1$.

\subsection{Conformal Blocks and Numerical Approach}

As in \cite{Zam}, the conformal block for the 4-pt function
$\la O_A(0)O_A(z)O_B(1)O_B(\infty)\lb$ can be expressed in the following way
 ($h$ is the dimension of intermediate primary):
\be\label{eq:Zamo}
F^{h_A,h_B}_h(z)=(16q)^{h-\frac{c-1}{24}}z^{\frac{c-1}{24}-2h_A}(1-z)^{\frac{c-1}{24}-h_A-h_B}
\cdot(\theta_3(q))^{\frac{c-1}{2}-8(h_A+h_B)}\cdot H^{h_A,h_B}_h(q),
\ee
where
\be
q=e^{i\pi\tau}=e^{-\pi\frac{K(1-z)}{K(z)}},
\ee
and
\be
\theta_3(q)=\sum_{n\in Z}q^{n^2}=\prod_{m=1}^\infty (1-q^{2m})(1+q^{2m-1})^2.
\ee
It is useful to note $K(0)=\frac{\pi}{2}$ and when $z$ is small we have
$K(1-z)\simeq \frac{1}{2}\log\frac{16}{z}$. We can express $z$ in terms of the theta functions
as $z=\left(\frac{\theta_2(q)}{\theta_3(q)}\right)^4$.

The function  $H^{h_A,h_B}_h(q)$ is found by solving the recursion relation in \cite{Zam}, whose
numerical prescription was formulated in \cite{Kaplan} (see appendix \ref{ap:rec} of this paper). It is expanded as
\be
H^{h_A,h_B}_h(q)=1+\sum_{n=1}^\infty c_n q^{2n}.  \label{defcn}
\ee

It is important to note that $H^{h_A,h_B}_h(q)$ has the symmetry:
\be
H^{h_A,h_B}_h(q)= H^{h_B,h_A}_h(q).  \label{symh}
\ee
This follows from the relation (\ref{ooab}) for the full 4-pt function is true for each conformal block. This relation is expected because it just flips $A$ with $B$ and does not change the structure of channel.

One useful limit we take is $q\to 1$. Accordingly $z$ approaches to $z=1$.
We write this as $z=1-\ep$ with $\ep\to 0$. Then we have in this limit:
\ba
&& q\equiv e^{-\delta},\ \ \delta\simeq \frac{\pi^2}{\log(16/\ep)}\to 0,\no
&& \theta_3(q)\sim \delta^{-1/2}. \label{qonel}
\ea

\subsection{Simplest Example of $H(q)$: Vacuum Primary}

If we consider the trivial limit of vacuum primaries $h_A=h_B=h=0$, obviously we have
$F^{h_A=h_B=0}_{h=0}(z)=1$. This fixes the form of $H(q)$ as follows
\ba
H^{0,0}_{0}(q)&=&(16q)^{\frac{c-1}{24}}\cdot z^{-\frac{c-1}{24}}
\cdot (1-z)^{-\frac{c-1}{24}}\cdot (\theta_3(q))^{-\frac{c-1}{2}}\no
&=&(16q)^{\frac{c-1}{24}}\cdot\left(\theta_2(q)\theta_3(q)\theta_4(q)\right)^{-\fr{c-1}{6}}\no
&=&q^{\frac{c-1}{24}}\cdot \eta(\tau)^{-\fr{c-1}{2}}\no
&=&\frac{1}{\left[\prod_{n=1}^\infty(1-q^{2n})\right]^{-\fr{c-1}{2}}}.
\ea
When $q$ is small (or equally $z$ is small), this is expanded as follows:
\ba
&&H^{0,0}_0(q)=\left(\prod_{n}(1-q^{2n})\right)^{-\frac{c-1}{2}} \no
&&= 1+\frac{1}{2}(c-1)q^2+\frac{1}{8}(c^2-1)q^4+\frac{1}{48}(c^3+3c^2-c-3)q^6+... \label{hqexp}
\ea

On the other hand, if we take the limit $z=1-\ep$ with $\ep\to 0$, we find
\be
H^{0,0}_{0}(q)\sim \delta^{\frac{c-1}{4}}\cdot \ep^{-\frac{c-1}{24}},
\ee
where $\sim$ means the approximation up to a constant factor.

By using the approximation formula of a summation (based on the saddle point approximation)
(we assume $A>0$):
\be
\sum_{n=0}^{\infty} n^\ap e^{A\s{n}}e^{-2n\delta}
\sim \delta^{-2\ap-\fr{3}{2}}e^{\frac{A^2}{8\delta}}. \label{saddlesum}
\ee
When $A=0$, we have
\be
\sum_{n=0}^{\infty} n^\ap e^{-2n\delta}\sim \delta^{-\ap-1}.  \label{sadn}
\ee

From (\ref{saddlesum}), we find the Cardy formula-like behavior of the coefficient $c_n$ defined in (\ref{defcn}) when $n\gg c$:
\be
c_n\simeq \beta\cdot n^\ap\cdot e^{A\s{n}}, \label{cnfor}
\ee
for a certain constant $\beta$ which we are not interested in.
Here $A$ and $\ap$ are given by
\ba
 A=\pi\sqrt{\frac{c-1}{3}}, \label{vaca} \ \ \  \ap=-\frac{c}{8}-\frac{5}{8}.
\ea

\subsection{Behaviors of $H(q)$}

Now we would like to examine the properties of $H(q)$ for general $h_A$ and $h_B$.
We focus on the vacuum conformal block $h=0$ as that is relevant for our later calculations of
Renyi entropy. First we can analytically calculate the coefficient $c_n$ assuming the large $c$ limit from the recursion relation:
\be
c_n\simeq \frac{1}{n!}\cdot \left(\frac{c}{2}\right)^n\cdot \left[\pa{1-\fr{32}{c}h_A}\pa{1-\fr{32}{c}h_B}\right]^{n}. \label{ksukif}
\ee
However note that here we ignored the lower powers of $c$ and this approximation is only sensible for $n\ll c$.

\begin{figure}[h]
 \begin{center}
  \includegraphics[width=7cm]{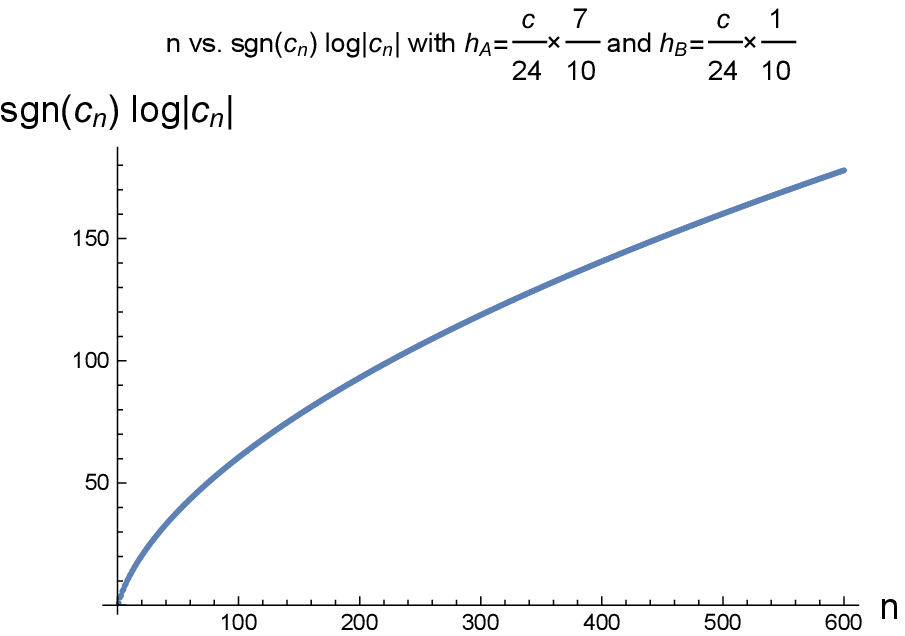}
  \includegraphics[width=7cm]{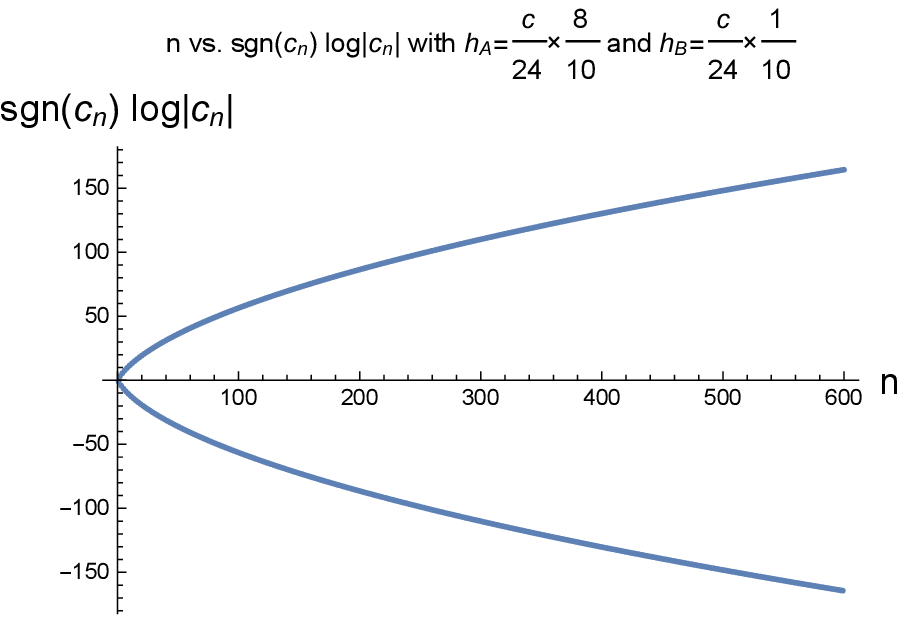}
 \end{center}
 \caption{The sign behaviors of coefficients $c_n$. Actually, we plotted
 sign$(c_n)\cdot \log|c_n|$ against $n$ for $n\leq 600$. The left plot is for
 $(h_A,h_B)=\frac{c}{24}(0.7,0.1)$, which is in the region (iii) and $c_n$ is always positive. The right one is for $(h_A,h_B)=\frac{c}{24}(0.8,0.1)$, which is in the region (ii) and has
 alternating signs.}
 \label{fig:sign}
\end{figure}

\begin{figure}[h]
 \begin{center}
  \includegraphics[width=8cm]{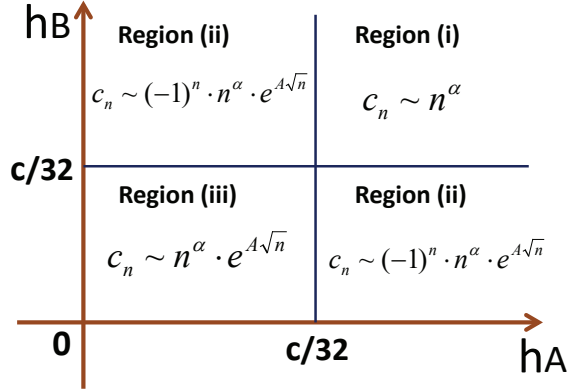}
 \end{center}
 \caption{The sketch of behaviors of $c_n$ for various values of $(h_A,h_B)$.}
 \label{fig:phase1}
\end{figure}

From this expression we find the following behavior of signs of $c_n$ (refer to Fig.\ref{fig:sign}):
\ba
(i) && h_A,h_B>\frac{c}{32} \hspace{5.2cm} : \ \ c_n= |c_n|, \label{sga} \\
(ii)&&  h_A>\frac{c}{32},\ h_B<\frac{c}{32} \ \mbox{or} \ h_A<\frac{c}{32},\ h_B>\frac{c}{32} :
\ \ c_n= (-1)^n|c_n|,   \label{sgb} \\
(iii) && h_A,\ h_B<\frac{c}{32} \hspace{5.1cm}  : \ \ c_n= |c_n|.  \label{sgc}
\ea
We summarized these behavior in Fig.\ref{fig:phase1}. Note that $c_n$ is invariant under the exchange of $h_A$ and $h_B$ as follows from the symmetry
(\ref{symh}).

\begin{figure}[h]
 \begin{center}
  \includegraphics[width=7cm]{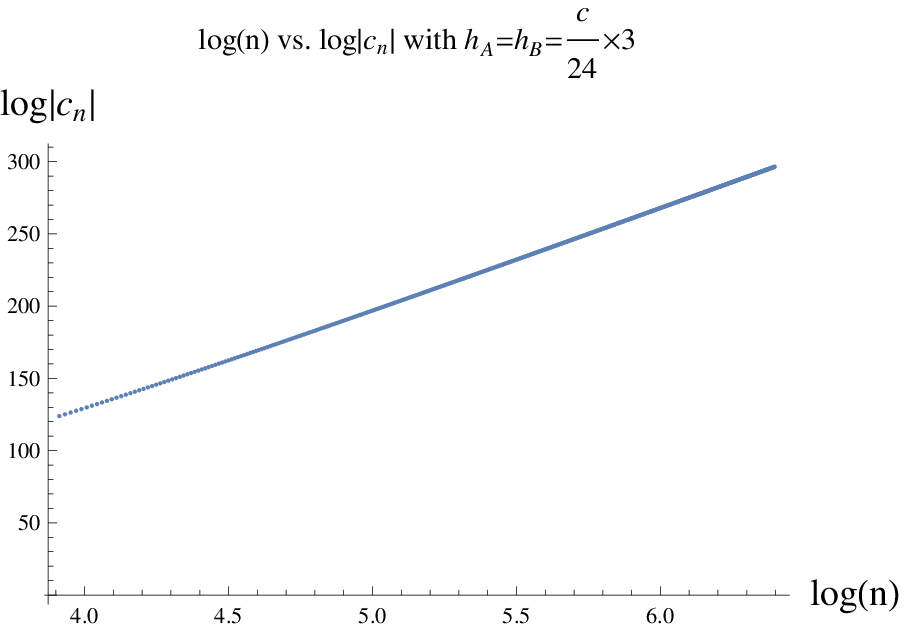}
    \includegraphics[width=7cm]{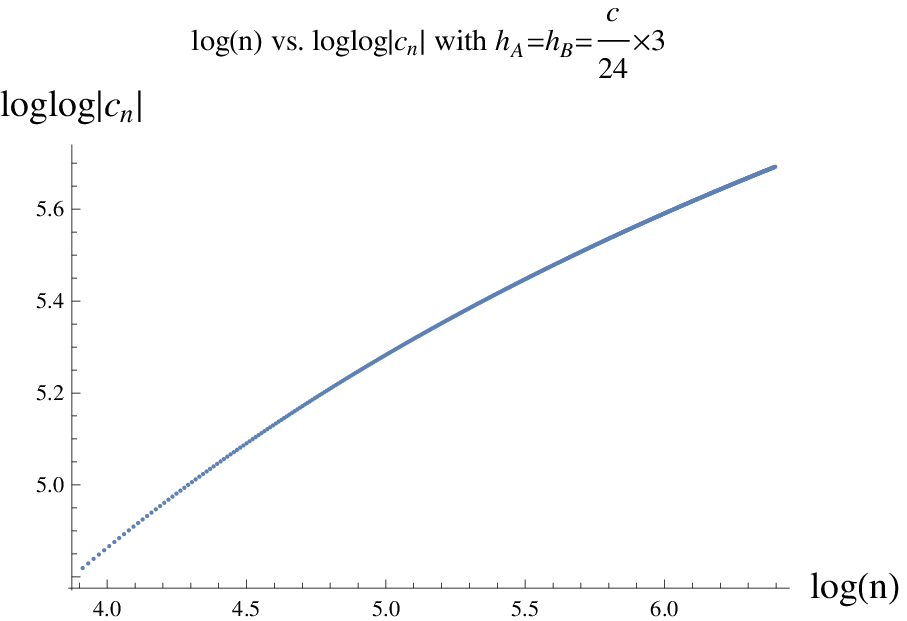}
 \end{center}
 \vspace{2mm}
  \begin{center}
  \includegraphics[width=7cm]{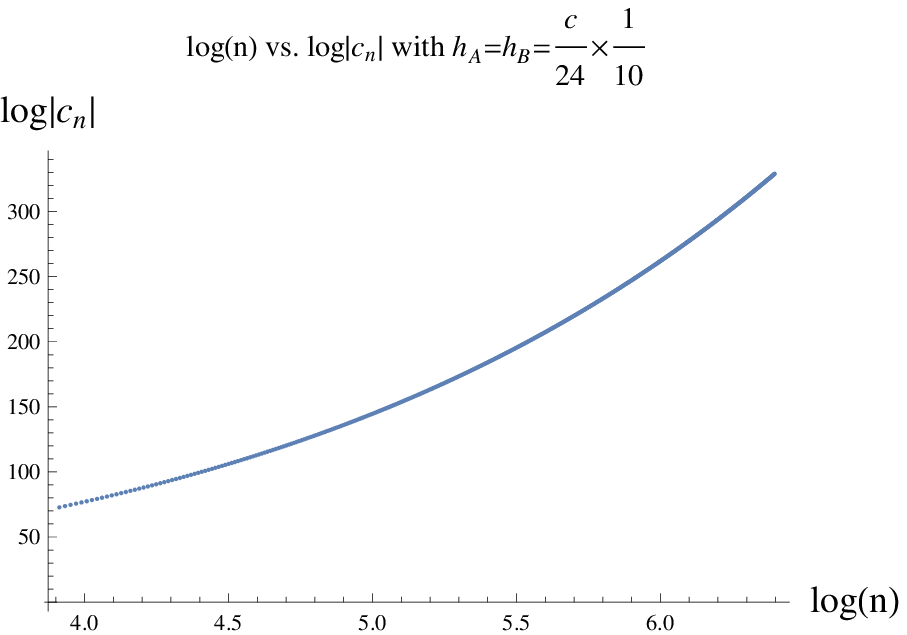}
    \includegraphics[width=7cm]{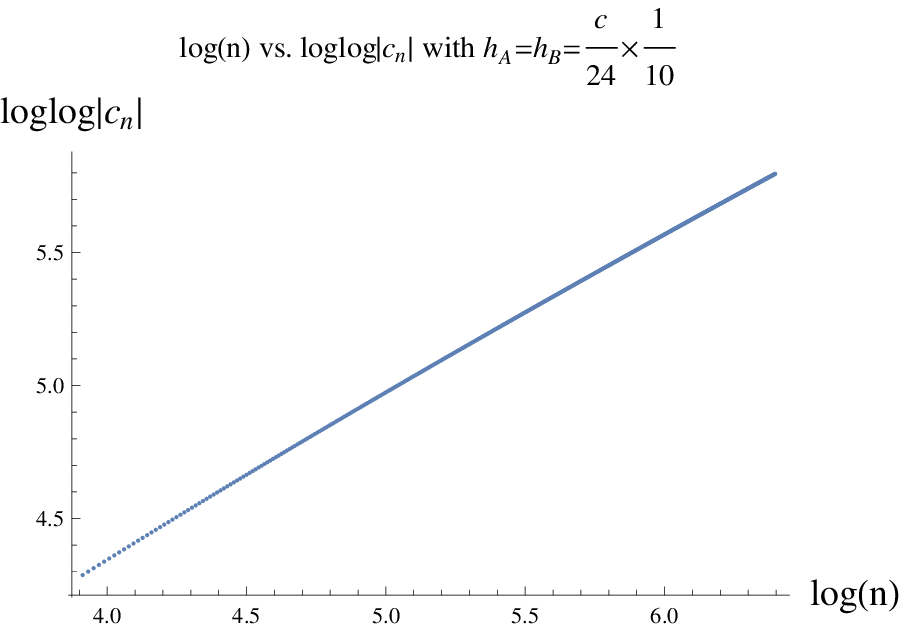}
 \end{center}
 \caption{The plots of $c_n$.  The upper two plots are for $h_A=h_B=\frac{c}{8}$ i.e. the region (i), where $c_n$ grows polynomially. The lower two plots are for $h_A=h_B=\frac{c}{240}$, i.e.
 the region (i), where $c_n$ grows exponentially.}
 \label{fig:cn}
\end{figure}

\begin{figure}[ttt]
\begin{center}
  \includegraphics[width=7cm]{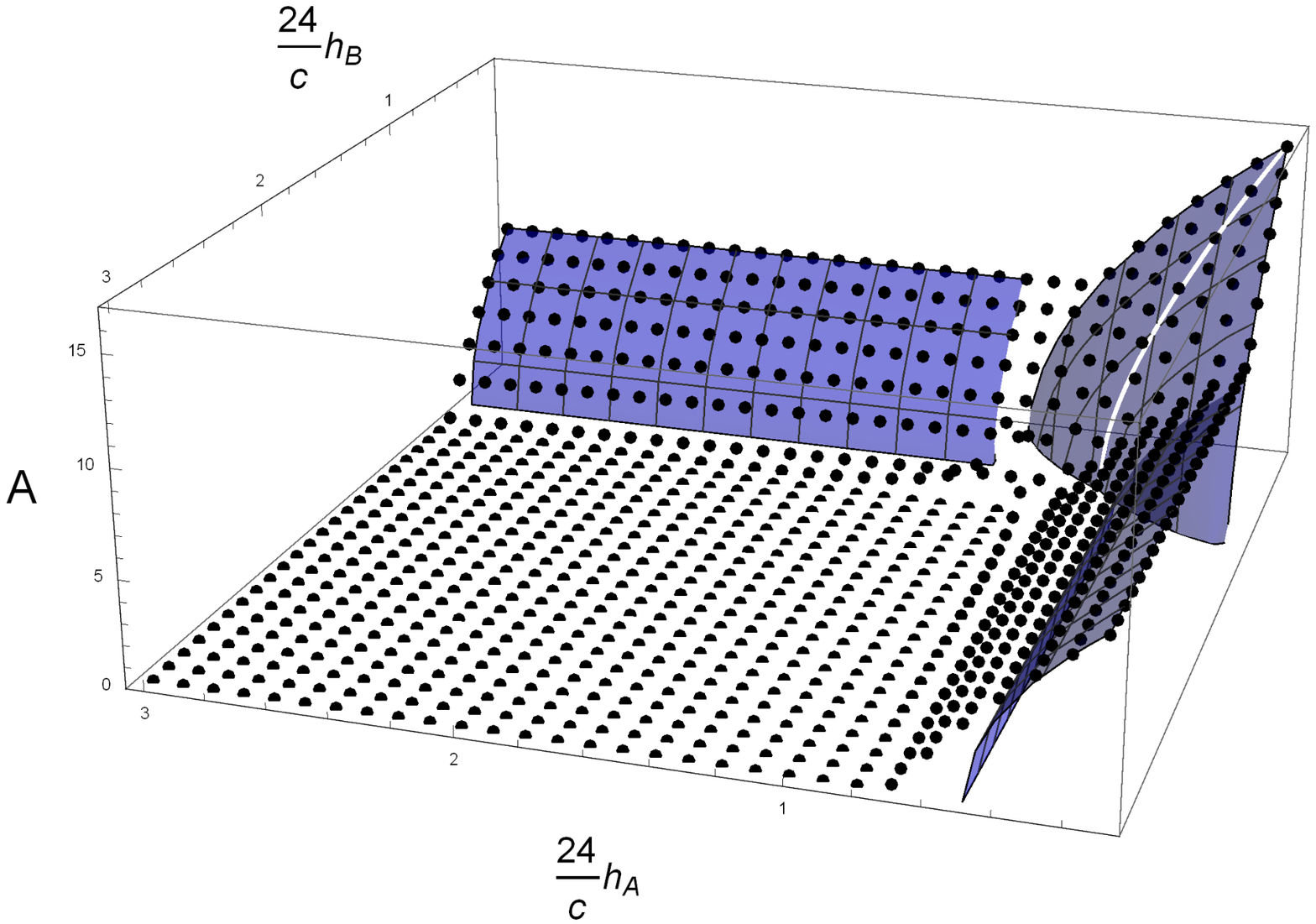}
   \includegraphics[width=7cm]{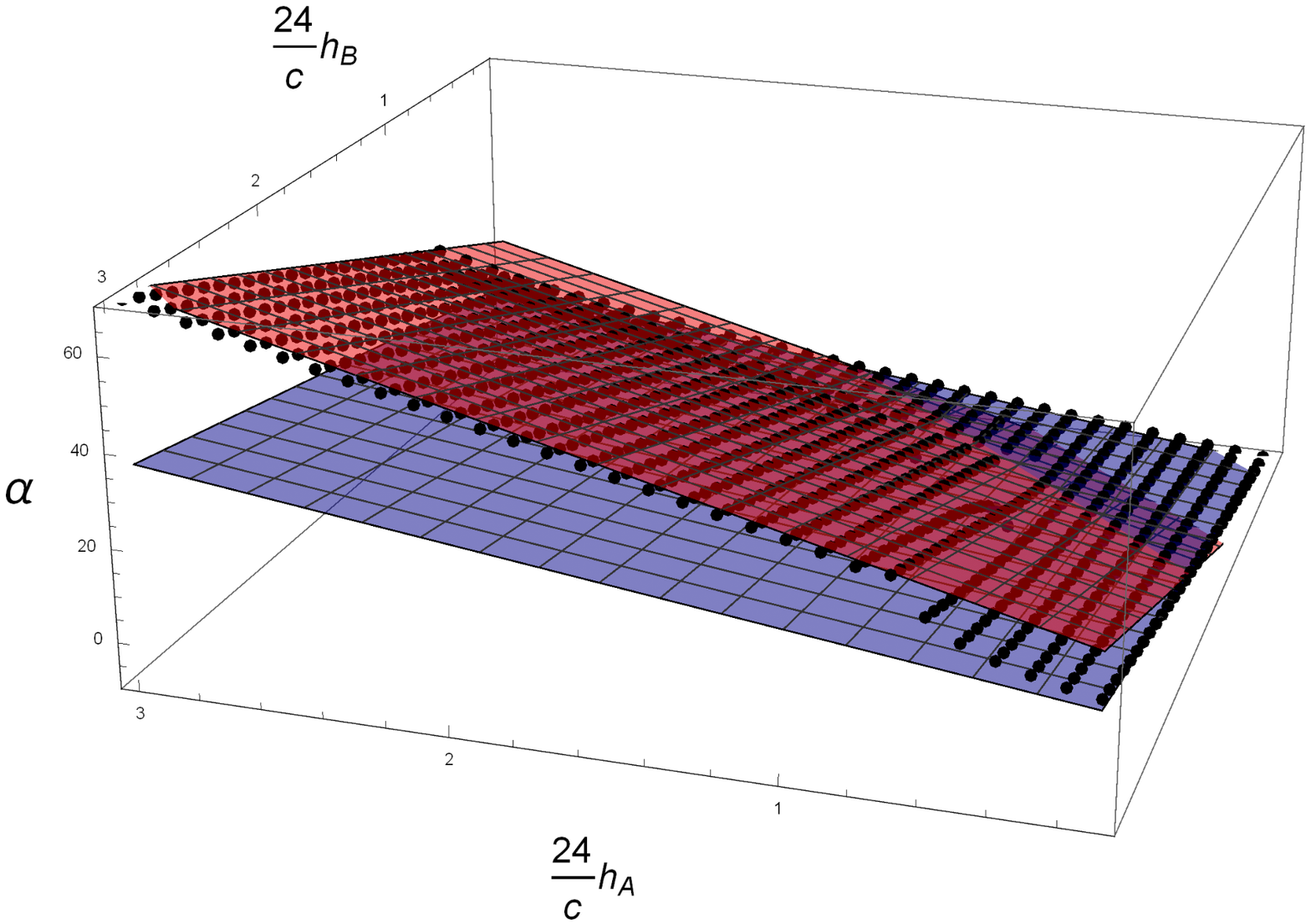}
\end{center}
 \caption{The plot of the values of $A$ (left) and $\ap$ (right) for various values of
 $h_A$ and $h_B$. The ranges are $0<h_A,h_B<\frac{c}{8}$. The black dots are the numerical values of $A$ and $\ap$. The blue surfaces are analytical predictions from the HHLL approximation. The red surface describes (\ref{numap}).
 We set $c=100.01$.}
 \label{fig:pows}
\end{figure}

Now we perform numerical computations of $c_n$ employing the computer program made in the paper
\cite{Kaplan} (for a short summary, see also appendix B of the present paper).
First, our numerical calculations of $c_n$ for various values of $(h_A,h_B)$
precisely show the above behaviors of signs, even for the regions $n\gg c$.
By fitting the numerical result for $c_n$ (refer to the plots Fig.\ref{fig:cn}) in the
Cardy formula-like form
\be
|c_n|\simeq \beta\cdot n^\ap\cdot e^{A\s{n}}, \label{cnforr}
\ee
and we evaluated the values of $A$ and $\ap$ for various $(h_A,h_B)$ and plotted in Fig.\ref{fig:pows}.

Interestingly we observe a clear transition of the values of $A$ and $\ap$
along the lines $h_A=\frac{c}{32}$ and $h_B=\frac{c}{32}$ as can be seen from the plots in Fig.\ref{fig:pows}. This behavior is
summarized as follows\footnote{In the paper \cite{Kaplan}, it was argued that the power law like behavior $c_n\propto n^s$ in the limit $n\to \infty$ can be observed for HHLL conformal block. Our analysis shows that this power law behavior (i.e. $A=0$) can only be seen for the region (i) and not in the region (ii) and (iii).}
\ba
&&(i)\ \  A\simeq 0,\no
&& (ii)\ \  A\simeq \pi\s{\frac{c}{12}}\cdot a_2(h_A,h_B)\ \ , \no
&& (iii)\ \  A\simeq \pi\s{\frac{c}{3}}\cdot a_3(h_A,h_B) \ \ .
\ea
Here the functions $a_2$ and $a_3$ are smooth monotonic functions bounded as
$|a_2|\leq 1$ and  $|a_3|\leq 1$. We have
\ba
&&\  a_2(0,h_B)=a_2(h_A,0)=a_3(0,0)=1, \label{xat} \\
&&\  a_3(c/32,0)=a_3(0,c/32)=\frac{1}{2},  \label{yat}\\
&&\  a_2(c/32,h_B)=a_2(h_A,c/32)=a_3(c/32,c/32)=0.
\ea
Note that $a_3(0,0)=1$ in (\ref{xat}) following from (\ref{vaca}).
As we will show in the next subsections, we can evaluate $A$ and $\ap$ by using the heavy-heavy-light-light (HHLL) approximation of conformal blocks assuming $h_A\equiv h_L\ll c$ and
$h_B\equiv h_H=O(c)$. This leads to
\ba
&& a_2(h_L,h_H)\simeq \s{1-\fr{48}{c}h_L}, \label{ya}\\
&& a_3(h_L,h_H)\simeq \s{1-\fr{24}{c}h_H-\fr{24}{c}h_L\s{1-\fr{24}{c}h_H}},  \label{xa}
\ea
Indeed, we find $a_3(h_L,c/32)\simeq \frac{1}{2}$, which justifies (\ref{yat}) and shows that $A$ is continuous at the border between (ii) and (iii). The value of $\ap$ is found as
\be
\ap^{HHLL}_{2,3}(h_L,h_H)=2(h_L+h_H)-\frac{c}{8}-\frac{5}{8},   \label{xp}
\ee
which can be applicable to both $(ii)$ and $(iii)$ if $h_L$ is small enough.

On the other hand, we would like to note that the result in the region $(i)$, $\alpha$ is
well fitted with the numerical data by the formula first considered in \cite{Kaplan}:
\be
\ap_1(h_A,h_B)=4(h_A+h_B)-\fr{c}{4}-\fr{9}{4}. \label{numap}
\ee
Note that the doubled coefficient of $h_{A}+h_B$ compared with (\ref{xp}) can be
understood from the factor two difference of the power of $\delta$ between the
$A>0$ formula (\ref{saddlesum}) and $A=0$ formula (\ref{sadn}).

\subsection{HHLL Approximation}

When one of $h_A$ and $h_B$ is light and another is heavy, we can apply the heavy-heavy-light-light (HHLL) approximation of conformal blocks \cite{FKW,FKWt}.
We take $h_A\equiv h_L\ll c$ and $h_B\equiv h_H=O(c)$.

In our case we have (we assume $h=0$)
\be
F^{h_H,h_L}_{h=0}(z)\simeq \left(\frac{\s{1-\fr{24}{c}h_H}}{1-(1-z)^{\s{1-\fr{24}{c}h_H}}}\right)^{2h_L}
\cdot (1-z)^{-h_L\pa{1-\s{1-\fr{24}{c}h_H}}}. \label{hhllap}
\ee
Thus in the $z\to 1$ limit: $z=1-\ep$, we get
\be
F^{h_H,h_L}_{h=0}(z)\simeq \ep^{-h_L\left(1-\s{1-\fr{24}{c}h_H}\right)}.
\ee
In terms of the function $H^{h_H,h_L}_{h=0}(q)$, this behavior is described as
\be
H^{h_H,h_L}_{h=0}(q)\sim \ep^{-\left(\frac{c-1}{24}\right)+h_H+h_L\s{1-\fr{24}{c}h_H}}
\cdot \delta^{\frac{c-1}{4}-4(h_H+h_L)}.\label{xx}
\ee
Now we focus on the region $(iii)$ i.e. $h_H<c/32$  so that $c_n$ is positive definite.
By using the formula (\ref{saddlesum}), we obtain the estimation of $A$ and $\ap$
given by (\ref{xa}) and (\ref{xp}).

\subsection{$q\to i$ limit}

When we consider the region $(ii)$ i.e. $h_H>c/32$, we need to worry about the alternating
signs in (\ref{sgb}), as $c_n$ behaves like $c_n\sim (-1)^n \cdot n^\ap\cdot e^{A\s{n}}$.
In such a case, it is not straightforward to obtain a formula like (\ref{saddlesum}).

To cancel the signs, we consider another limit of $q\to i$. This is obtained by
$z\to 0$ limit with a monodoromy transformation around $z=1$. Explicitly we have
\be
q_{mo}=e^{-\pi\frac{K(1-z)}{K(z)+2iK(1-z)}},\ \  \ z=\ti{\ep}\to 0.
\ee
This behaves like (we define $\ti{\delta}=\frac{\pi^2}{\log(16/\ti{\ep})}$):
\be
q_{mo}\simeq e^{\pi i\tau_m}\simeq i\cdot e^{-\fr{\ti{\delta}}{4}},\ \ \ \ \ \tau_{mo}=\frac{1}{2-i\ti{\delta}/\pi}\equiv\frac{1}{2-\tau}, \label{qwz}
\ee

From the HHLL approximation (\ref{hhllap}),
we obtain the following behavior for the modular limit $q\to i$
\be
H^{h_H,h_L}_{h=0}(q_{mo})\sim \ti{\ep}^{-\left(\frac{c-1}{24}\right)+2h_L}\cdot\ti{\delta}^{\frac{c-1}{4}-4(h_H+h_L)}.\label{yy}
\ee
By using the formula (\ref{saddlesum}), we can read off from this behavior the advertised values of $A$ and $\ap$ given by (\ref{ya}) and (\ref{xp}).

Actually, this limit $q\to i$ exactly corresponds to the one we need to calculate the Renyi entropy described in (\ref{monoz}) i.e. $q_{mo}=q(z_{mo})$. In this relation, we can identify
$\ti{\ep}=-\frac{2i\ep}{t}$.

\section{Evaluation of Renyi Entropy}

Now we are in a position to study the Renyi entropy computed by the formula
(\ref{eq:Fform}) based on our previous results for the vacuum conformal block.
First note that to calculate the $n$-th Renyi entropy for a large central charge CFT
(holographic CFT) with a central charge $c_{CFT}$
by the replica method, we consider a CFT with the central charge $c=n\cdot c_{CFT}$, defined by taking $n$ copies of the original CFT. Then we take
\be
h_A=h_{\sigma_n}=\frac{c_{CFT}}{24}\pa{n-\fr{1}{n}}, \ \ \ h_B=nh_O.
\ee
The growth of  Renyi entropy is symmetric under the exchange of $h_A$ and $h_B$ as
\ba
\Delta S^{(n)}_A &=&\frac{1}{1-n}\log \left[z^{2h_A}\bar{z}^{2h_A}F^{h_A,h_B}_{0}(z)F^{h_A,h_B}_{0}(\bar{z})\right]_{z=z_{mo},
\bar{z}\to 0}\no
&=&\frac{1}{1-n}\log \left[z^{2h_B}\bar{z}^{2h_A}F^{h_B,h_A}_{0}(z)F^{h_B,h_A}_{0}(\bar{z})\right]_{z=z_{mo},\bar{z}\to 0}\no.
\ea
Since we act the monodromy transformation only for $z$ and not for $\bar{z}$,
in the limit (\ref{monoz}) we have $q\to i$ as in (\ref{qwz}) and $\bar{q}\simeq \frac{\bar{z}}{16}\to 0$.

Thus we can simply (\ref{eq:Fform}) as follows
\be
\Delta S^{(n)}_A \simeq \frac{1}{1-n}\log \left[(z_{mo})^{\frac{n\cdot c_{CFT}-1}{24}}\cdot H(q_{mo})\right], \label{www}
\ee
where we neglect subleading terms and keep such terms grow as $\sim |\log\ti{\ep}|\sim \log(\fr{t}{\e})$ in the end.\footnote{In other words, we neglect terms $\sim \log\log \fr{t}{\e}$.}

In general, we find that $\Delta S^{(n)}_A$ grows logarithmically under time evolutions.
Therefore, below, we are interested in the coefficient of the log $t$ term, denoted by
$B(n,h_O)$:
\be
\Delta S^{(n)}_A \simeq B(n,h_O)\cdot \log\frac{t}{\ep}.  \label{bcoef}
\ee

\begin{figure}[ttt]
\begin{center}
  \includegraphics[width=7cm]{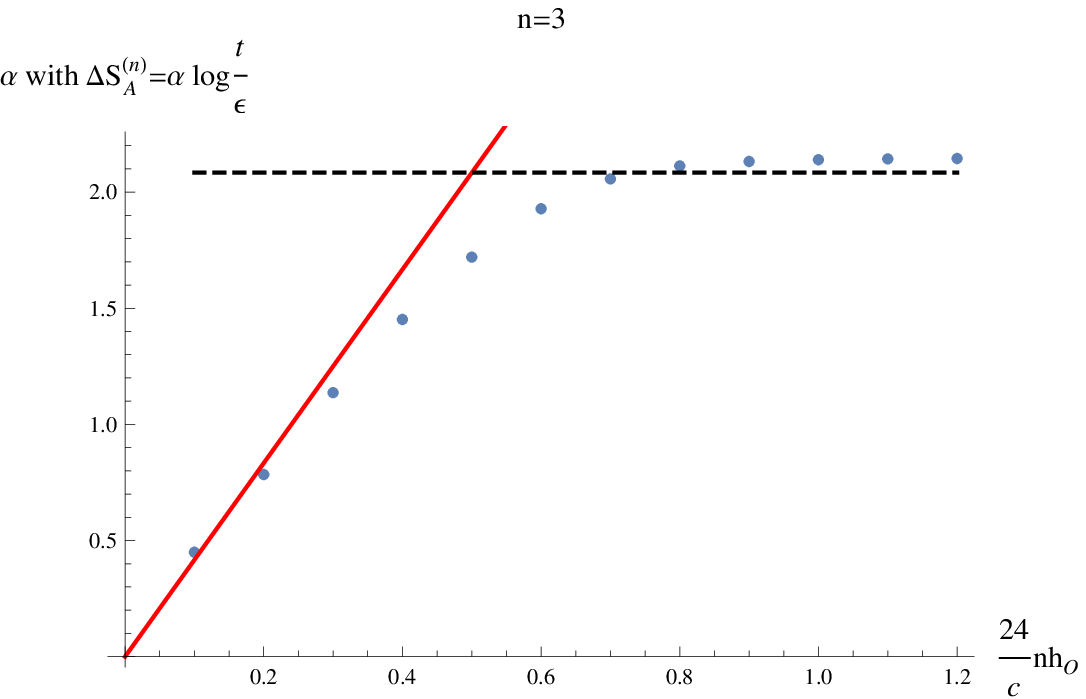}
   \includegraphics[width=7cm]{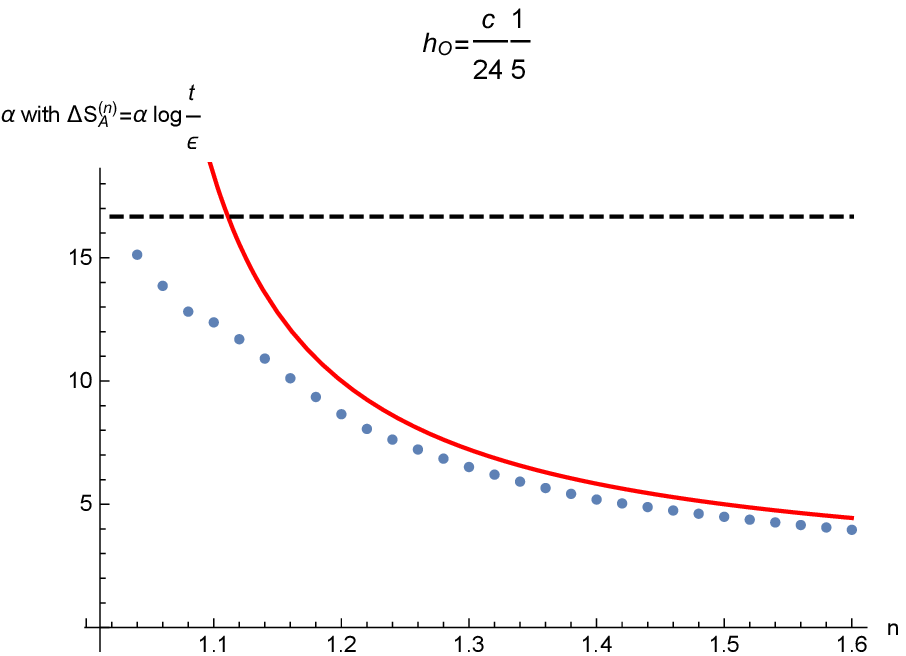}
\end{center}
 \caption{The plots of the coefficient $B(n,h_O)$ of the logarithmic growth of entanglement
 entropy. The left is the case of the third Renyi entropy $n=3$ for various values of
 $h_O$ (the vertical dotted line (A) in Fig.\ref{fig:phase2}). The red line describes the formula (\ref{capee}) and the dotted horizontal line corresponds to the formula (\ref{reone}). The right is the case of $h_O=\frac{c}{120}=\frac{n\cdot c_{CFT}}{120}$
 for various values of $n$
 (the horizontal dotted line (B) in Fig.\ref{fig:phase2}). The dotted horizontal line and red curve correspond to  the formula (\ref{holee}) and  (\ref{capee}), respectively. }
 \label{fig:crossp}
\end{figure}

The behaviors of $B(n,h_O)$ are summarized in Fig.\ref{fig:phase2}. Our numerical results of $B(n,h_O)$ are plotted in Fig.\ref{fig:crossp} as we will explain below.

\subsection{HHLL Approximation}

If we apply the HHLL approximation with $h_A\ll c\sim h_B$, we obtain from (\ref{yy})
\be
\Delta S^{(n)}_A\simeq \frac{2h_A}{n-1}\log \frac{t}{\ep}.
\ee
If we choose the light operator $O_A$ to be the twist operator $\sigma_n$ in the limit $n\to 1$, then we reproduce the formula (\ref{holee}). If we choose the light operator $O_A$ to be $O$, namely the operator for the excitation, then we reproduce another formula (\ref{capee}). Note that these approximated results are continuous at $h_{A,B}=c/32$. The regions where we can apply this HHLL approximation is depicted in Fig.\ref{fig:phase2} as the red and green regions.

\subsection{Region $(i)$}

In the region $(i)$ defined by (\ref{sga}), we simply have
\be
B(n,h_O)=\frac{nc_{CFT}}{24(n-1)},\label{reone}
\ee
namely the advertised formula (\ref{kee}). This is derived as follows. First note that $H(q_{mo})$ does not give any leading divergence
which contributes to $O(\log t)$ entropy, remembering the behavior $c_n\sim n^\ap$ (i.e. no $e^{A\s{n}}$ factor). Then the formula (\ref{www}) with $nc_{CFT}\gg 1$ leads to the formula (\ref{reone}). Indeed this behavior is confirmed in the left plot in Fig.\ref{fig:crossp}.

At the special value $n=2$ and $h_O=\frac{c_{CFT}}{32}$, the conformal block coincides with a torus partition function \cite{He,HaSC,HMPZ}. By using this fact we can evaluate $\Delta S^{(2)}_A$ as computed in \cite{Pawel}, which indeed agrees with (\ref{reone}).\footnote{We are very much grateful to Pawel Caputa and Tomonori Ugajin for pointing out this to us.}

\subsection{Region $(ii)$}

In the region (ii) defined by (\ref{sgb}), we can apply the formula (\ref{saddlesum}) in $q\to i$ limit as we have explained in the previous section. Thus we can calculate the coefficient $B$ in (\ref{bcoef}) in terms of $A$:
\be
B(n,h_O)=\frac{1}{n-1}\left(\frac{nc_{CFT}}{24}-\frac{A^2}{2\pi^2}\right)
=\frac{nc_{CFT}}{24(n-1)}\left(1-(a_2)^2\right).
\ee
This function interpolates the HHLL approximated result and the formula (\ref{reone}) in the region $(i)$. Examples of the plots of $B(n,h_O)$ in this region can be found from the both two plots in Fig.\ref{fig:crossp}.

\subsection{Region $(iii)$}

The region (iii), defined by (\ref{sgc}) also includes parameter spaces where we can apply the HHLL approximation. Also when $n=2$ and $h_O=\frac{c_{CFT}}{32}$, we should reproduce the behavior (\ref{reone}). Even though we do not have any analytical formula, our numerical results show that the function $B(n,h_O)$ monotonically interpolates these boundary values, as depicted in the right plot
of Fig.\ref{fig:crossp}.

\subsection{Comment on OTOC}
In this paper, we focus on entropy, but the vacuum block we derived can be also used to evaluate OTOC.  As in \cite{Roberts:2014ifa}, we can express the late time behavior of OTOC in holographic CFTs as
\begin{equation}
\fr{\ave{O_A (t)O_B O_A(t) O_B}_\beta}{\ave{O_A O_A}_\beta \ave{O_B O_B}_\beta}\simeq \abs{z_{mo}}^{4h_A}
|F^{h_A,h_B}_{0}(z_{mo})|^2.
\end{equation}
The cross ratio $z_{mo}$ is given by
\bea
z_{mo}\simeq -e^{-\frac{2\pi (t-x)}{\beta}}\epsilon^*_{12}\epsilon_{34},\qquad
\bar{z}_{mo}\simeq -e^{-\frac{2\pi (t+x)}{\beta}}\epsilon^*_{12}\epsilon_{34},\label{CRbeta}
\eea
where $\epsilon_{ij}=\ii\left(e^{\frac{2\pi \ii}{\beta}\epsilon_i}-e^{\frac{2\pi \ii}{\beta}\epsilon_j}\right)$ and $x$ is the separation between operators that keeps the ratio $\bar{z}/z$ fixed. This technique is very similar to that used to calculate entropy. Therefore  we can apply our result to calculating OTOC straightforwardly and it leads to the late time behavior of OTOC for any two operators, in particular two heavy operators. The result is as follows.
\begin{equation}
\fr{\ave{O_A (t)O_B O_A(t) O_B}_\beta}{\ave{O_A O_A}_\beta \ave{O_B O_B}_\beta} \simeq
\ex{-\fr{c-1}{12}\fr{\pi t}{\beta}}, \ \ \ \ \ \ \mbox{if } h_A, h_B>\fr{c}{32}.
\end{equation}
and in the heavy-light limit, we can reproduce the results in \cite{Roberts:2014ifa, Perlmutter:2016pkf}. Moreover we can conclude that the behaviors of OTOC show the late exponential decay for any operator at late time. This exponential decay cannot be seen in
non-chaotic CFTs, where the OTOC approaches non-zero constant\cite{Caputa:2016tgt, Gu:2016hoy} or decays polynomially \cite{Caputa:2017rkm}. This may suggest that this late time behavior can also be used as a criterion of chaotic nature of a given quantum field theory, in addition to the existing arguments on the Lyapunov exponent \cite{Roberts:2014ifa,MSS,FKC}. It is also intriguing to note that the above exponential decay behavior of OTOC is directly related to the logarithmic growth of Renyi entanglement entropy (\ref{kee}).

\vspace{1cm}

{\bf \large Acknowledgments}

We thank Arpan Bhattacharrya, Kanato Goto, Yasuaki Hikida, Fabio Novaes, Shinsei Ryu, and Tomonori Ugajin for useful discussions, and in particular Pawel Caputa and Jared Kaplan for reading the draft of this paper and giving us valuable comments. TT is supported by the Simons Foundation through the ``It from Qubit'' collaboration. TT is supported by JSPS Grant-in-Aid for Scientific Research (A) No.16H02182. TT is also supported by World Premier International Research Center Initiative (WPI Initiative) from the Japan Ministry of Education, Culture, Sports, Science and Technology (MEXT). YK and TT are very grateful to the workshop ``Holography and Dynamics'' (YITP-X-17-06), held in Yukawa Institute for Theoretical Physics, Kyoto University where results of this paper were presented.

\appendix

\section{Convention of 4-pt Function}

Here we summarize our conventions on 4-pt functions and conformal blocks.
Consider two kinds of primary operators $O_A$ with the dimension $(h_A,\bar{h}_A)$ and
$O_B$ with the dimension $(h_B,\bar{h}_B)$. The full expression of 4-pt function,
written as $\la O_A(w_1)O_A(w_2)O_B(w_3)O_B(w_4)\lb$, takes the following form
\ba
&& \la O_A(w_1)O_A(w_2)O_B(w_3)O_B(w_4)\lb  \no
&& =\left|w_{12}^{-\frac{4}{3}h_A+\frac{2}{3}h_B}w_{13}^{-\frac{1}{3}(h_A+h_B)}
w_{14}^{-\frac{1}{3}(h_A+h_B)}w_{23}^{-\frac{1}{3}(h_A+h_B)}
w_{24}^{-\frac{1}{3}(h_A+h_B)}w_{34}^{-\frac{4}{3}h_B+\frac{2}{3}h_A}\right|^2
 W(z,\bar{z}), \no \label{fpt}
\ea
where $z=\frac{(w_1-w_2)(w_3-w_4)}{(w_1-w_3)(w_2-w_4)}$ is the cross ratio.

We define the (normalized) 4-pt function $\la O_A(0)O_A(z)O_B(1)O_B(\infty)\lb$ by taking the
limit $(w_1,w_2,w_3,w_4)\to(0,z,1,\infty)$ and by absorbing the divergence as follows:
\be
\la O_A(w_1)O_A(w_2)O_B(w_3)O_B(w_4)\lb\to \left|(\infty)^{-2h_B}\right|^2\cdot \la O_A(0)O_A(z)O_B(1)O_B(\infty)\lb.  \label{fpta}
\ee
In other words we find the relation
\ba
 G(z,\bar{z})&\equiv&  \la O_A(0)O_A(z)O_B(1)O_B(\infty)\lb \no
&=&\left|z^{-\frac{4}{3}h_A+\frac{2}{3}h_B}(1-z)^{-\frac{1}{3}(h_A+h_B)}\right|^2W(z,\bar{z}).
\ea

On the other hand if we take the limit $(w_1,w_2,w_3,w_4)\to(1,\infty,0,z)$, we can define
$\la O_A(1)O_A(\infty)O_B(0)O_B(z)\lb$ as follows:
\be
\la O_A(w_1)O_A(w_2)O_B(w_3)O_B(w_4)\lb\to \left|(\infty)^{-2h_A}\right|^2\cdot \la O_A(1)O_A(\infty)O_B(0)O_B(z)\lb. \label{fptb}
\ee
By comparing (\ref{fpta}) and (\ref{fptb}) based on the expression (\ref{fpt}), we find the
relation
\be
\la O_A(1)O_A(\infty)O_B(0)O_B(z)\lb\cdot |z|^{4h_B}=\la O_A(0)O_A(z)O_B(1)O_B(\infty)\lb \cdot
|z|^{4h_A}.  \label{ooab}
\ee
This relation is very natural because it is just an exchange of two $O_A$s with two $O_B$s.

The (normalized) 4-pt function can be written as the summation over all conformal blocks:
\ba
\la O_A(0)O_A(z)O_B(1)O_B(\infty)\lb=\sum_{p}C_{AAp}C_{BBp}
F^{h_A,h_B}_{h_p}(z)F^{h_A,h_B}_{\bar{h}_p}(\bar{z}),\label{ff}
\ea
where $h_p$ is the conformal dimension of the intermediate primary state.

\section{Recursion Relations}\label{ap:rec}

In this Appendix, we will review Zamolodchikov's recursion relation. In our case $h_1=h_2=h_A$ and $h_3=h_4=h_B$, the Virasoro conformal block can be expressed as
\be
F^{h_A,h_B}_h(z)=(16q)^{h-\frac{c-1}{24}}z^{\frac{c-1}{24}-2h_A}(1-z)^{\frac{c-1}{24}-h_A-h_B}
\cdot(\theta_3(q))^{\frac{c-1}{2}-8(h_A+h_B)}\cdot H^{h_A,h_B}_h(q)
\ee
and $H^{h_A,h_B}_h(q)$ is given by the following recursion relation,
\ba
H^{h_A,h_B}_h(q)=1+\sum^\infty_{\subs{m=1,n=1}{ mn\in \mbox{\scriptsize even}}}\frac{q^{mn}R_{m,n}}{h-h_{m,n}}H^{h_A,h_B}_{h_{m,n}+mn}(q),
\label{rech}
\ea
where $h_{m,n}$ is a zero of the Kac determinant and
\ba
R_{m,n}=2\cdot\frac{\left(\prod_{p,q}\lambda_{p,q}\right)^2}{\prod'_{k,l}\lambda_{k,l}}
\cdot \prod_{p,q}(2\lambda_A-\lambda_{p,q})(2\lambda_B-\lambda_{p,q}).
\label{rechh}
\ea
Here the integers $p,q,k,l$ are defined as
\ba
&&p=-m+1,-m+3,\cdots,m-3,m-1,\no
&&q=-n+1,-n+3,\cdots,n-3,n-1,\no
&&k=-m+1,-m+2,\cdots,m,\no
&&l=-n+1,-n+2,\cdots,n.
\ea
The product $\prod'_{k,l}$ in (\ref{rechh}) means that we exclude $(k,l)=(0,0)$ and $(m,n)$.
We also defined
\ba
&&c=1+\pa{b+\fr{1}{b}}^2,\no
&&h_{A,B}=\fr{c-1}{24}-\lambda_{A,B}^2,\no
&&\lambda_{p,q}=\fr{1}{2}\pa{\fr{p}{b}+qb}.
\ea
We expand $H^{h_A,h_B}_h(q)$ as
\be
H^{h_A,h_B}_h(q)=1+\sum_{k=1}^\infty c_k(h) q^{2k}.
\ee
In the same way as (\ref{rech}), we can also calculate the coefficients $c_k(h)$ recursively by the following relation,
\begin{equation}
	c_k(h) = \sum_{i=1}^k \sum_{\subs{m=1, n=1}{mn=2i}} \frac{R_{m,n}}{h - h_{m,n}} c_{k-i}(h_{m,n}+mn),
\end{equation}
where the sum is took over $m,n=1,2,3,\cdots$ with $mn$ held fixed, i.e. the sum $\sum_{\subs{m=1, n=1}{mn=4}}$ means taking sum over $(m,n)=(1,4), (2,2), (4,1)$. The coefficient $c_{k}(h_{m,n}+mn)$ can be also calculated recursively by
\begin{equation}
	c_k(h_{m,n}+mn) = \sum_{i=1}^k \sum_{\subs{\m=1, \n=1}{\m\n=2i}} \frac{R_{\m,\n}}{h_{m,n}+mn - h_{\m,\n}} c_{k-i}(h_{\m,\n}+\m\n),
\end{equation}
where the starting values of this recursion formula are $c_0(h_{m,n}+mn)=1$. Note that in this paper, we describe $c_k(0)$ as $c_k$.

\end{document}